\documentclass[11pt]{article}

\usepackage[latin1]{inputenc}
\usepackage{amsmath}
\usepackage{amsfonts}
\usepackage{amssymb}
\usepackage{xspace}

\renewcommand\iff{if and only if\xspace}
\renewcommand\th{\theta}
\def\l{\left}  \def\r{\right}

\newcommand\LB[1]{\label{#1}} 
\newcommand\BE[2]{\begin{#1} #2 \end{#1}}

\newcommand\EQ[2]{\BE{equation}{\LB{#1} #2}}

\newcommand\proof{\noindent {\sc Proof:}\qquad}
\newcommand\qed{\hfill$\quad${\rule{3mm}{3mm}}\medskip\\}

\newcommand{\eq}{equation\xspace}

\def\a{\alpha}
\def\b{\beta}

\newcommand\om{\omega}

\newcommand \sg{\sigma}

\newcommand\Th{\Theta}

\newcommand{\cL}{\mathcal{L}}

\newcommand\dl{\delta}

\newcommand{\bbR}{\blackboard{R}}
\newcommand{\blackboard}[1]{\mathbb#1}
\newcommand{\HS}{H^2(\Re z>0)}
\newcommand{\iy}{\infty}

\def\const{\mathrm{const}}

\title{%
{\bf Inverse Problems for the Heat Equation with Memory
 }
 }

\author{
Sergei A. Avdonin\thanks{University of Alaska, Fairbanks,  AK 99775-6660, USA, {\tt s.avdonin@alaska.edu }}
	\thanks{The research
		of S.A. Avdonin was supported in part by the National Science Foundation,
		grant DMS 1411564, and by the Ministry of Education and Science of Republic of Kazakhstan under the grant no. 4290/GF4.}
\and
{S.A. Ivanov\thanks{Russian Academy of Sciences, St. Petersburg, Russia,
{\tt sergei.a.ivanov@mail.ru}}
\thanks{The research
of S.A. Ivanov was supported in part by the
Russian Foundation of Basic Research,  grant 14-01-00349a.} }
\and
{ Jun Min Wang
\thanks { School of Mathematics and Statistics,
Beijing Institute of Technology, Beijing 100081, P.R. China, {\tt jmwang@bit.edu.cn}}
\thanks{The research
of J.M. Wang was supported in part by the National Natural Science Foundation of China,
grant 61673061.}
 }
}

\date{}

\begin{document}
\date{}
\maketitle

\begin{abstract}
\noindent
We study inverse boundary problems  for  one dimensional linear
integro-differential equation of the Gurtin--Pipkin type with the
Dirichlet-to-Neumann map as the inverse data. Under  natural conditions on the
kernel of the integral operator, we give the explicit formula for the
solution of the problem with the observation on the semiaxis $t>0.$
 For the observation on finite time interval, we prove the
uniqueness result, which is similar to the local Borg--Marchenko theorem
for the Schr\"odinger equation.


\end{abstract}

\vskip1cm

\section{\LB{intro_etc} Introduction and the main results}

\subsection{\LB{intro_etc1}{ Gurtin--Pipkin type equations}}

It is known that the classical heat \eq has a non-physical property, namely the infinite speed of
propagation of singularities.
Based on the modified Fourier law,  Gurtin and Pipkin \cite{GuPip}
introduced  a model of heat transfer with a finite propagation
speed\footnote{The propagation of singularities was studied in \cite{JNR}, \cite{Ivel}}.
In the present paper we consider a form of this model described by the linear integro-differential \eq
\begin{equation}\label{1a}
\theta_t(x,t)=
\int_0^t k(t-s) \theta_{xx} (x,s)d s,  \quad  t>0, \  x\in(0,L), \ L \leq \iy,
\end{equation}
the initial condition  $\, \theta(\cdot,0)=0\,$ and the boundary conditions
\begin{equation}\label{1b}
\th(0,t)=f(t), \ \th(L,t)=0.
\end{equation}
The latter condition will be omitted in the case $L = \iy.$ Conditions on the kernel $k$
will be discussed in Section 1.2.

Another form of the Gurtin--Pipkin model (an isotropic viscoelastic model) is described by the integro-differential equation of the second order in time:
\begin{equation}\LB{2GP2}
u_{tt}(x,t)=a u_{xx}+\int_0^t k(t-s) u_{xx}(x,s)\,ds,\ a>0,\ x\in(0,L),  \ t>0.
\end{equation}
The following form of the heat \eq with memory,
\EQ{s}{
u_{t}(x,t)=u_{xx}+\int_0^t k(t-s) u_{xx}(x,s)\,ds, \ x\in(0,L), \ t>0,
}
can also be found in the literature.

We will consider the equation \eqref{1a} because in this
form the integral term plays the most important role.
In some interpretations of the equation this means the absence of the  latent heat.
Note that the differentiation of \eqref{1a} with respect to $t$ leads to the \eq of the form \eqref{2GP2}.
The system described by \eqref{s}
has the infinite speed of  propagation of singularities
and it is closer, in this sense, to the heat \eq than \eqref{1a}.

Equation \eqref{1a} can be treated as a perturbed  wave \eq.
In the case $k(t)=\const=\a^2$, \eq \eqref{1a} is
in fact an integrated wave \eq. Indeed,
differentiating \eqref{1a} we obtain
$
u_{tt}=\a^2 u_{xx}.$
If $k(t)=e^{-bt},$
then differentiation gives a damped wave \eq
$u_{tt}=u_{xx}-bu_t.$
In the singular case $k(t)=\dl(t)$ the  \eq \eqref{1a} becomes the heat \eq.

\subsection{\LB{intro_etc2}{Well-posedness of initial boundary value problems}}

Well-posedness of initial boundary value problems for the Gurtin--Pipkin type equations was studied by many authors. In particular, regularity of the solutions in the Sobolev spaces was a topic of a series of papers of V.Vlasov and his coauthors, see, e.g.\cite{RSV} and also  \cite{I2013}.
For the case of a finite time interval regularity results for several forms
of Gurtin--Pipkin type equations can be found in the book
\cite{Luc}. Note that in these papers the kernel $k$ is assumed to be continuous and positive at the origin.

For the case of a finite space interval  the solution can be obtained using the Fourier approach (see the example in section 3.2).
The solution can also be constructed by the Laplace transform, see section \ref{s1}.

Now we state the assumption on the kernel that we use in the present paper.
In what follows we suppose that the Laplace transform of $k$ satisfies the condition
\EQ{K0}{
K(z)=\frac{a^2}z+O\l(\frac{1}{z^2}\r),\ a>0, \ \Re z>0.
}
Roughly speaking this means that $k'$ is bounded and $k(0)=a^2>0$.

\BE{remark}{ If $k(0)=0$ or $k$ has a singularity, the solution may not exist even in a weak sense.
In the example at the end of the paper we show that in the case $k(t)=t^2$ the \eq \eqref{1a} has
no solution in  Sobolev spaces.
}

\subsection{\LB{intro_etc3}{ The statement of the inverse problem}}

Let $\,T\le \iy\,$,  $\,f \in L^2(0,T)\,$ and $\,\th^f\,$ be a solution to the initial boundary value problem \eqref{1a}, \eqref{1b}.
We introduce  the response operator $r^T: L^2(0,T) \mapsto L^2_{loc}(0,T)$ with the domain
$\,\{f \in H^1(0,T),\, f(0)=0\}\,$ acting by the rule
\EQ R{
(r^Tf)(t) =\th_x^f(0,t), \
\  t \in (0,T).
}
Our inverse problem is formulated as follows: given the response operator $r^T,\,$ to
recover the kernel $k$ on the maximal possible interval $[0,T_0].$
(We will demonstrate that this interval is exactly  $[0,T].$)

Such a statement of inverse problem is standard  for the hyperbolic type equations, see, e.g  \cite{ABI}.
However, the known methods do not work in our situation. For example, the boundary control method successfully used in \cite{ABI} for recovering a matrix potential $Q(x)$ is
not applicable to the problem where unknown coefficient depends on time.

It is  convenient
to extend the response operator to distributions including the Dirac delta function.
Due to linearity of the Gurtin--Pipkin equation,
the response operator is completely determined by the response to the Dirac delta control,
as well as in the case of the wave type \eq \cite{ABI}.

\subsection{\LB{intro_etc4}{A brief survey of known results on inverse problems for equations with memory}}

There are many papers concerning inverse problems for partial differential equations with memory. However, almost all of them deal with
unknown sources or (in multidimensional cases)  with
unknown spatial part of the kernel $k(x,t)$ if it has a form $k(x,t)=h(t)p(x).$ Here we mention several papers devoted to inverse problems similar to ours.

In  \cite{BKK} the \eq of the second order with memory in $\bbR^3$ was studied:
$$
u_{tt}=\Delta u - k*u, \ \mbox{in $\bbR^3$}, \ u|_0=0,\ u_t|_0=\dl(x-x_0).
$$
The inverse data is the value of the scattering wave at $x_0$ for  $0<t<T$.
The authors demonstrate the uniquely stable identification of $k$ on the time interval $[0,T-|x_0|]$.

In \cite{JW} the inverse problem for the system describing by the \eq
$$
\b u_t= u_{xx} - m*u_{xx}, \ x\in (0,1),
$$
was studied. The properties of this system is close
to a parabolic type \eq.
The authors prove that by, for instance,  given $u(x_0,t), \,t>0,$
it is possible to recover the kernel $m$.

In \cite{APan,Pan1,Pan2} the inverse problem for the system \eqref{1a}, \eqref{1b} was studied
on a finite spatial interval $(0,L)$ with the help of the Fourier method. A linear algorithm
reconstructing the kernel $k$ from two boundary observation was developed. One of the observations corresponds to a nonzero initial condition.

\subsection{\LB{intro_main_res}{ Main results}}

In the present paper we study inverse problem for the system \eqref{1a}, \eqref{1b} on a finite spatial interval and on the
semi-axis with the help of the Laplace transform.
In the case of the infinite time of observation we obtain explicit formulas that allow recovering the Laplace transform of the  kernel $k.$

In the case of a finite time of observation we prove the uniqueness
of the solution to the inverse problem and obtain  the local uniqueness result
 similar to  the local Borg--Marchenko theorem for the Schr\"odinger \eq \cite{S,Be}. Our approach is based on the Laplace
 transform and uses some basic facts of the Hardy space theory.

 We recall now the local Borg--Marchenko theorem.
 The following uniqueness result was proved  in \cite{S} (see also a  very short proof in \cite{Be}).
 If two Weyl-Titchmarsh m-functions, $m_j(z), \, j=1,2,$ for two Schr\"odinger equations
 $$
 d^2/dx^2 \psi - q_j \psi = z \psi, \ x >0,
 $$
with some regular condition at $x=0$
 are exponentially close, that is,
 $$
 |m_1(z)-m_2(z)|\le C e^{-2\Im\sqrt za}, \  \Im \sqrt z >0,
 $$
 then $q_1 = q_2$ on $[0,a]$. This result may be considered as a local version of
 the celebrated Borg--Marchenko uniqueness theorem \cite{M}.

 In the present paper we prove a similar local uniqueness result for kernel of the heat equation with memory.

\section{\LB{s1} The stationary inverse problem, $T=\iy$}

The inverse data is $r^\iy,$ the observation of $\th_x(0,t)$ for all $t>0$.

We suppose that the solution to the IBVP \eqref{1a}, \eqref{1b} does not grow too fast in order to be able to apply the
Laplace transform. This assumption can be justified using the integral representation of the solution presented in \cite{Luc} (section 1.2, formula (1.10)) and the condition \eqref{K0}.

We apply the Laplace  transform to \eqref{1a}, \eqref{1b}, denote
the images by capitals and
 obtain the family of  ODEs depending on $z$ as a parameter:
\EQ{new1}{
z\Th(x,z)=K(z)\Th_{xx}(x,z), \ x\in(0,L), \ \Th(0,z)=F(z).
}

For every $z$ this differential (in $x$) equation has constant coefficients. We set
$\,
\om(z)=\sqrt{z/K(z)}\,
$
(the main branch) and consider separately the cases of a finite and the
infinite interval $[0,L]$.

\subsection{The case $L=\iy$}

\BE{theorem}{\LB{stat} Let $k$  satisfy \eqref{K0} and $L=\iy$.
Then the kernel $k$ can be uniquely recovered from $r^\iy$.
}
\proof
The  solution to \eqref{new1} which do not
increasing exponentially in the right half $z$-plane is
$$
\Th(x,z)=F(z)e^{-\om(z)x}.
$$
Then the Laplace transform of the response $\th_x(0,t)$ is
\EQ{R(z)}{
R(z)=\Th_x(0,z)=-F(z)\om(z)=-F(z)\sqrt{z/K(z)}.
}
Evidently, we can find $K(z)$ via the data $R$ and the given $F$.
\qed

\subsection{The case of a finite interval, $L<\iy$ }

\BE{theorem}{\LB{stat1} Let $k$  satisfy \eqref{K0} and $L<\iy$.
Then the kernel $k$ can be uniquely recovered from $r^\iy$.
}
\proof
In this case the solution $\Th$ satisfies the \eq \eqref{new1} and  zero boundary condition at
$x=L$. Thus, we have the problem
$$
\Th_{xx}(x,z)=\om^2\Th(x,z), \ \Th(0,z)=F(z), \ \Th(L,z)=0.
$$
First, taking into account only the boundary condition at $x=0,$ we obtain
$$
\Th(x,z)=F(z)\cosh[ \om(z)x]+ \Phi(z)\sinh [ \om(z)x].
$$
The  boundary condition at $x=L$ implies
$$
\Th(L,z)=F(z)\cosh[ \om(z)L]+ \Phi(z)\sinh [ \om(z)L]=0.
$$
Then for $\om(z)L \ne \pi n\,$, $n \in \mathbb{Z},$
$$
\Phi(z)=-\frac{F(z)\cos[ \om(z)L]}{\sin [ \om(z)L]}.
$$
From the other hand,
$$
R(z)=\Th_x(0,z)=\om(z)\Phi(z)=-\om(z) \frac{F(z)\cos[ \om(z)L]}{\sin [ \om(z)L]}.
$$
It is possible now to recover $\om(z)$,
then $K(z)$ and  $k(t)$.
\qed

\section{\LB{s2} Non-stationary inverse problem,  $T<\iy$}
  \subsection{The local uniqueness}

First, we consider the case  $L=\iy.$
Note that, by the condition  \eqref{K0},  the function $K$ can have only finite number of zeros in the right half plane.
For simplicity we suppose now that $K(z)$ has no zeros there:
$K(z)\ne 0, \ \Re z>0$.
This takes place, in particular, if $k$ is a non-increasing function.

Now the inverse data is
$
r(t)=\th_x(0,t), \, t<T,\,$
corresponding to the control
$
f(t)=t^+.
$
Then $F=\frac1{z^2}$ and by \eqref{R(z)} and the assumption on $K$ we obtain that
$$
 R(z)=- \frac 1{az}+O\left(\frac{1}{z^2}\right), \ \Re z >0,
$$
and $R$ is analytical in the right half plane.
In fact we need  $R$ to be in the right Hardy space.

\BE{theorem}{$k_1(t)=k_2(t)$ , $0\le t \le T,$ \iff
$R_1^T=R_2^T$.
}
\proof The part \textbf{`only if'}.
We have
$$
K_1(z)-K_2(z)=\int_T^\iy e^{-zt} (k_1(t)-k_2(t))dt.
$$
This means that
\EQ{K12}{
e^{zT}( K_1(z)-K_2(z)) \in \HS.
}
From the explicit expression for the response operator we find
$$
R_1(z)-R_2(z)=
\frac{\sqrt{z}}{z^2\sqrt{K_1K_2}}\l(\sqrt{K_2}-\sqrt{K_1}\r)
$$
or
\EQ{R12}{
R_1(z)-R_2(z)=
\frac{\sqrt{z}}{z^2\sqrt{K_1K_2}\l(\sqrt{K_2}+\sqrt{K_1}\r)}\l(K_2-K_1\r).
}
By the main assumption \eqref{K0} the factor
$$
\frac{\sqrt{z}}{z^2\sqrt{K_1K_2}\l(\sqrt{K_2}+\sqrt{K_1}\r)}
$$
is asymptotically equal to $1/a^3$.

Then \eqref{K12} implies
$$
e^{zT}( R_1-R_2) \in \HS
$$
what is equivalent to $r_1(t)=r_2(t)$ , $0\le t \le T$.

Part \textbf{`if'}.
The equality
$$
R_1^T=R_2^T,
$$
can be written as
$$
e^{zT}( R_1-R_2) \in \HS.
$$
From \eqref{R12} we see that
$$
e^{zT}( K_1-K_2) \in \HS.
$$
This means that $k_1(t)=k_2(t)$ for $0\le t \le T$.
\qed

Now we derive an interesting  integral relation between $R^T$ and $K$.
Evidently,  we can not find the whole $R$
from the inverse data, but we know its projection $R^T$ onto
$$
K_T=\HS\ominus e^{-zT}\HS.
$$

We write $R^T$ in the terms of $R(z)=R^\iy(z)$.
For simplicity we write the integral in the inverse Laplace transform
over the imaginary axis (not over $i\bbR+\sg$).
$$
\cL[\chi_{[0,T]}r(t)](z)
=\int_0^\iy e^{-tz} \chi_{[0,T]}r^\iy(t)dt
$$
$$
=\int_0^T dt\, e^{-tz}\frac1{2\pi i}\int_{-i\iy}^{i\iy}dp\,e^{pt}R(p)
\overset{p=iy}=\frac1{2\pi}\int_{-\iy}^{\iy}dy\, R(iy)\int_0^T dt\, e^{t(iy-z)}
$$
$$
=\frac1{2\pi} \int_{-\iy}^{\iy} \frac{e^{T(iy-z)}-1}{iy-z} R(iy)\,\,dy.
$$
Thus the problem is to understand what information about $k$  can be extracted from the \eq
$$
R^T(z)=-\frac1{2\pi} \int_{-\iy}^{\iy} \frac{e^{T(iy-z)}-1}{iy-z}\frac1{y^2}
 \sqrt{\frac{iy}{K(iy)}}\,\,dy
$$
with known $R^T(z)$.

 Let now $L<\iy$, and $x \in (0,L).$  Since the system \eqref{1a}, \eqref{1b}
has a finite speed of the wave propagation equal to $a,$ we have the same uniqueness
theorem as for the semi-axis.

\BE{theorem}{$k_1(t)=k_2(t)$ , $0\le t \le T,$ \iff
	$R_1^{T_*}=R_2^{T_*},$ where $T_*=Ta \leq L.$ }

 \subsection{Example}
Let us give an example of a `non Sobolev' solutions  of
\eqref{1a}, \eqref{1b} in the case where $k(0)=0$
(and \eqref{K0} is not true )
Take a smooth kernel vanishing at the origin, say, $k(t)=t$.
Its Laplace image is $K(z)=1/z^2$.

We will find the solution as a series in sine functions
(the eigenfunction of the operator $d^2/dx^2$ with the Dirichlet boundary conditions)
$$
\th(x,t)=\sum_1^\iy \th_n(t)\sin nx.
$$
Let the initial data be
$$
\th(0,x)=\sum_1^\iy \xi_n\sin nx.
$$
For the Laplace image of $u_n$ we have  \cite{IE}
$$
\Th_n(z)=\frac{\xi_n}{z+n^2K(z)}=\frac{z^2\xi_n}{z^3+n^2},
$$
and
The pre-image $\th_n(t)$  of $\Th_n(z)$ is
$$
\cL^{-1}\frac{z^3\xi_n}{z^4+n^2}
=\frac{\xi_n}3\l(
e^{-n^{2/3}t}+e^{\exp(-\pi i/3)n^{2/3}t}+e^{\exp(\pi i/3)n^{2/3}t}
\r).
$$

Hence $n^{-P}\th_n(t)$ is in $\ell^2$ for fixed
$t$ only if the coefficients decrease exponentially as $n^Pexp(-n^{2/3}t/2)$.
It is easy to see that the same conclusion is correct for any kernel $t^\a$
with $-1<\a$, $\a\ne 0$.

\end{document}